%% file: paper.tex
\pdfoutput=1
\documentclass[10pt,journal,cspaper,compsoc]{IEEEtran}
\usepackage{amssymb}
\usepackage{graphicx}
\usepackage{amsmath}
\newcommand{\ignore}[1]{}
\newif\ifarxivversion
 \arxivversiontrue  

\begin{document}
%
\title{Gradient-Domain Processing for Large EM Image Stacks}
%
%
%
%

\author{Michael~Kazhdan, Randal~Burns, Bobby~Kasthuri, Jeff~Lichtman,\\Jacob~Vogelstein, and~Joshua~Vogelstein
\IEEEcompsocitemizethanks{%
\IEEEcompsocthanksitem M. Kazhdan and R. Burns are with the Department of Computer Science, Johns Hopkins University.
\IEEEcompsocthanksitem B. Kasthuri and J. Lichtman  are with the Department of Molecular and Cellular Biology, Harvard University.%
\IEEEcompsocthanksitem J. Vogelstein is with the Johns Hopkins Applied Physics Laboratory.%
\IEEEcompsocthanksitem J. Vogelstein is with the Department of Statistical Science, Duke University.}%
\thanks{}}

%
%

\markboth{Gradient-Domain Processing for Large EM Image Stacks
}%
{Gradient-Domain Processing for Large EM Image Stacks}
%


\IEEEcompsoctitleabstractindextext{%
\begin{abstract}
\input{abstract}
\end{abstract}

}

\maketitle

\IEEEdisplaynotcompsoctitleabstractindextext

%
\IEEEpeerreviewmaketitle

\section{Introduction}
\label{s:intro}
\input{intro}

\section{Related Work}
\label{s:related}
\input{related}

\section{Voxel Processing}
\label{s:gradient-domain}
\input{gradient_domain}

\section{Solving the Linear System}
\label{s:implementation}
\input{implementation}

\section{Results}
\label{s:results}
\input{results}

\section{Discussion}
\label{s:discussion}
\input{discussion}

\section{Conclusion}
\label{s:conclusion}
\input{conclusion}

\ignore
{
\appendices
\section{Proof of the First Zonklar Equation}
Appendix one text goes here.

\section{}
Appendix two text goes here.Appendix two text goes here.Appendix two text goes here.Appendix two text goes here.Appendix two text goes here.Appendix two text goes here.Appendix two text goes here.Appendix two text goes here.Appendix two text goes here.Appendix two text goes here.Appendix two text goes here.Appendix two text goes here.Appendix two text goes here.Appendix two text goes here.Appendix two text goes here.Appendix two text goes here.Appendix two text goes here.Appendix two text goes here.Appendix two text goes here.Appendix two text goes here.Appendix two text goes here.Appendix two text goes here.Appendix two text goes here.Appendix two text goes here.Appendix two text goes here.
}

\ifCLASSOPTIONcompsoc
  \section*{Acknowledgments}
\else
  \section*{Acknowledgment}
\fi


\ifCLASSOPTIONcaptionsoff
  \newpage
\fi



%

\bibliographystyle{IEEEtran}
\bibliography{paper}

\ignore
{

}
%

\ignore
{
\begin{IEEEbiography}{Michael Shell}
Biography text here.
\end{IEEEbiography}

\begin{IEEEbiographynophoto}{John Doe}
Biography text here.Biography text here.Biography text here.Biography text here.Biography text here.Biography text here.Biography text here.Biography text here.Biography text here.Biography text here.Biography text here.Biography text here.Biography text here.Biography text here.Biography text here.Biography text here.Biography text here.Biography text here.Biography text here.Biography text here.Biography text here.Biography text here.Biography text here.Biography text here.Biography text here.Biography text here.Biography text here.Biography text here.Biography text here.Biography text here.Biography text here.Biography text here.
\end{IEEEbiographynophoto}


\begin{IEEEbiographynophoto}{Jane Doe}
Biography text here.Biography text here.Biography text here.Biography text here.Biography text here.Biography text here.Biography text here.Biography text here.Biography text here.Biography text here.Biography text here.Biography text here.Biography text here.Biography text here.Biography text here.Biography text here.Biography text here.Biography text here.Biography text here.Biography text here.Biography text here.Biography text here.Biography text here.Biography text here.Biography text here.Biography text here.Biography text here.Biography text here.
\end{IEEEbiographynophoto}
}




\end{document}

%% file: abstract.tex
We propose a new gradient-domain technique for processing registered EM image stacks to remove the inter-image discontinuities while preserving intra-image detail. To this end, we process the image stack by first performing anisotropic diffusion to smooth the data along the slice axis and then solving a screened-Poisson equation within each slice to re-introduce the detail. The final image stack is both continuous across the slice axis (facilitating the tracking of information between slices) and maintains sharp details within each slice (supporting automatic feature detection). To support this editing, we describe the implementation of the first multigrid solver designed for efficient gradient domain processing of large, out-of-core, voxel grids.

%% file: intro.tex
Recent innovation and automation of electron microscopy sectioning has made it possible to obtain high-resolution image stacks capturing the relationships between cellular structures~\cite{Hayworth:MM:2006}. This, in turn, has motivated research in areas such as  connectomics~\cite{Anderson:BP:2009,Biswal:PNAS:2010,Bock:N:2011,Anderson:MV:2011} which aims to gain insight into neural function through the study of the connectivity network.

\begin{figure}[h]
\center
{
\ifarxivversion
\includegraphics[width=0.32\columnwidth,natwidth=496,natheight=496]{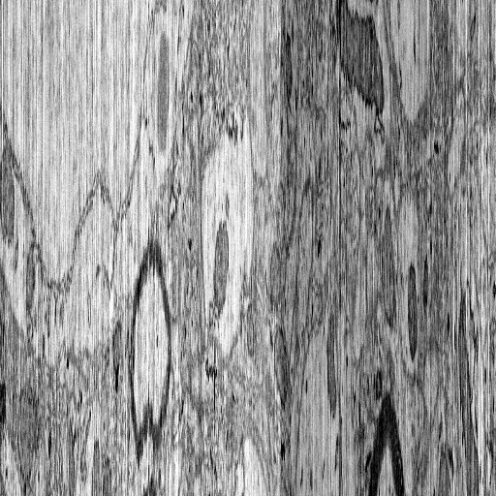}
\includegraphics[width=0.32\columnwidth,natwidth=496,natheight=496]{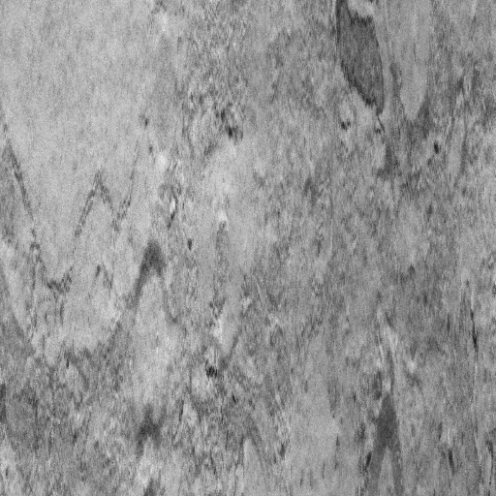}
\includegraphics[width=0.32\columnwidth,natwidth=496,natheight=496]{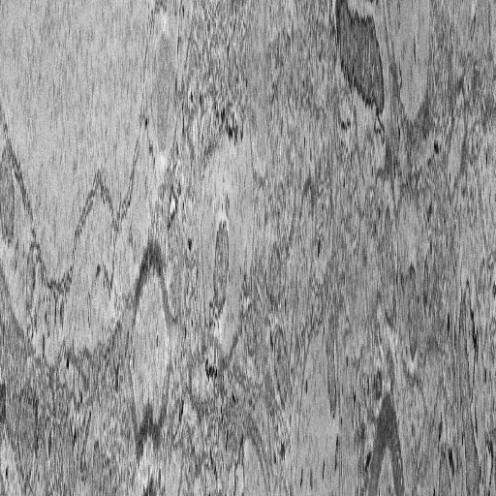}
\fi
}
\center
{
\ifarxivversion
\includegraphics[width=0.32\columnwidth,natwidth=496,natheight=496]{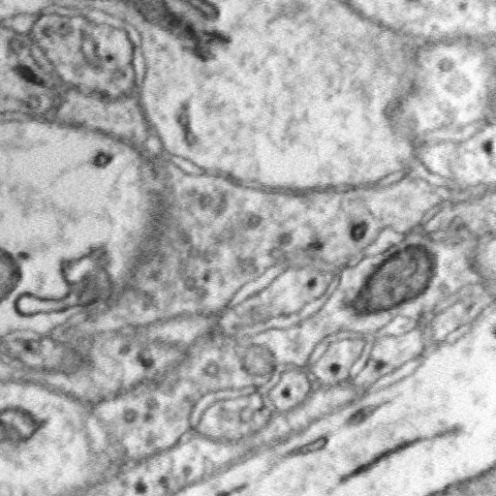}
\includegraphics[width=0.32\columnwidth,natwidth=496,natheight=496]{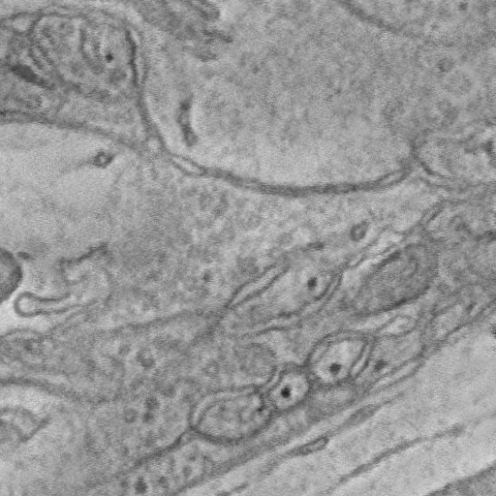}
\includegraphics[width=0.32\columnwidth,natwidth=496,natheight=496]{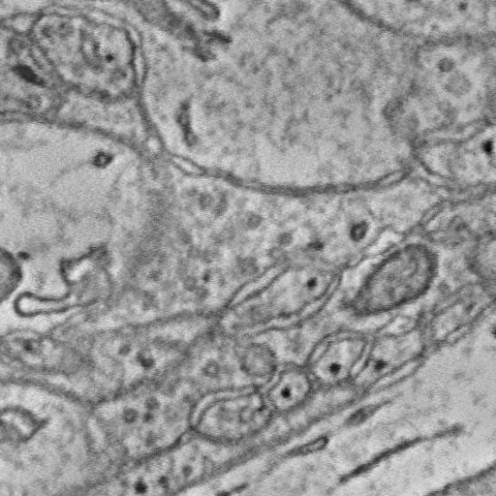}
\fi
}
\caption{
\label{f:data}
Cross-sections of an EM stack showing $zy$-slices through the data (top) and $xy$-slices through the data (bottom). The image on the left is taken from the original data, the image in the center is the result of the initial anisotropic diffusion step, and the image on the right is the subsequent solution of the screened-Poisson equation.
}
\end{figure}

While the technological advances in acquisition and registration have made it possible to acquire unprecedentedly large micron-resolution volumes, the acquisition process itself introduces undesirable artifacts in the data, complicating tasks of (semi-)automatic anatomy tracking. Specifically, since the individual slices in the stack are imaged independently, discontinuities often arise between successive slices due to variations in lighting, camera parameters, and the physical manner in which a slice is positioned on the slide. An example of these artifacts can be seen in Figure~\ref{f:data} (top-left), which shows an image of the same column taken from successive images in a stack (1850 images at a resolution of $21496\times25792$) imaging a mouse cortex~\cite{Kat11}. The visualization highlights both the local discontinuities (thin vertical stripes across the image) and global discontinuities (brighter band on the left of the image versus darker band on the right) that can arise due to the acquisition process.

In this work we propose a new gradient-domain technique for processing these anatomical volumes to remove the undesired artifacts. The processing consists of two phases. In the initial phase, we perform anisotropic diffusion across the slice axis to smooth out the discontinuities between these slices. As Figure~\ref{f:data} (center) shows, this has the desirable effect of removing the discontinuities (top) but it also smooths out the anatomical features within the slice (bottom). To address this, we perform a second step of gradient-domain processing on each slice independently, solving a screened-Poisson equation to generate a new voxel grid with low-frequency content taken from the anisotropically diffused grid and high-frequency content taken from the original data. As  Figure~\ref{f:data} (right) shows, this combines the best parts of both datasets -- like the anisotropically diffused grid, this solution does not exhibit discontinuities between slices (top), while simultaneously preserving the sharp detail present in the original data (bottom).

Our implementation of the gradient-domain processing is enabled by a new, parallel and out-of-core multigrid solver, designed to support a broad class of gradient domain techniques over large datasets. Using our solver, we are able to complete both phases of gradient-domain processing over the entire tera-voxel dataset in less than 60 hours, parallelized across 12 cores and using 45GB of RAM. 


%% file: related.tex
Over the last decade, gradient-domain approaches have gained prevalence in image
processing~\cite{Agrawal:ICCV:2007}. Examples include removal of light and shadow
effects~\cite{Horn:CGIP:1974,Finlayson:ECCV:2002}, reduction of dynamic
range~\cite{Fattal:SIGGRAPH:2002,Weyrich:SIGGRAPH:2007}, creation of intrinsic
images~\cite{Weiss:ICCV:2001}, image stitching~\cite{Perez:SIGGRAPH:2003,Agarwala:SIGGRAPH:2004,Levin:ECCV:2004},
removal of reflections~\cite{Agrawal:SIGGRAPH:2005}, and gradient-based sharpening~\cite{Bhat:ECCV:2008}.

The versatility of gradient-domain processing has led to the design of numerous methods
for solving the underlying Poisson problem in the context of large 2D images, including
adaptive~\cite{Agarwala:SIGGRAPH:2007}, out-of-core~\cite{Kazhdan:SIGGRAPH:2008}, and
distributed~\cite{Kazhdan:TOG:2010} solvers. However, to date, no solvers have been designed
for supporting analogous processing over large voxel grids.

This work presents both a new class of gradient-domain techniques, targeted at removing errors
specific to volumetric data, and a new solver supporting this processing over large
voxel grids.

In domains seeking to better understand the functioning of the human brain (e.g.~\cite{BrainInitiative})
these provide indespensible tools for working with the multi-terabyte datasets returned by new
scanners. In more traditional graphics applications, these extend established techniques for
processing 2D images to the processing of large video in a temporally coherent manner.

\ignore
{
Taking advantage of the human eye's sensitivity to local color variation, these approaches enable an image
processing paradigm in which the user formulate local constraints and the system automatically solves for the
image that is the best fit to the constraints.
Examples have included the removal of light and shadow effects by zeroing either small
gradients~\cite{Horn:CGIP:1974} or gradients crossing shadow boundaries~\cite{Finlayson:ECCV:2002}.
Remapping of dynamic range, both in images~\cite{Fattal:SIGGRAPH:2002} and in height
fields~\cite{Weyrich:SIGGRAPH:2007}, has been implemented through nonlinear attentuation of gradient
magnitudes.
Combining information from multiple images of the same scene, the median gradients have been used to derive
intrinsic images~\cite{Weiss:ICCV:2001} and remove reflections~\cite{Agrawal:SIGGRAPH:2005}.
And recently, methods for the {\em ex nihilo} construction of images from gradients has been used to
design new image authoring paradigms~\cite{McCann:SIGGRAPH:2008,Orzan:SIGGRAPH:2008}.
}

%% file: gradient_domain.tex

\subsection{Gradient Domain Formulation}
We begin by formulating the filtering of the signal at two successive steps of gradient-domain processing. We then show how these steps can be interpreted in the frequency domain.
\subsubsection{Anisotropic Diffusion}
First, we perform an anisotropic diffusion across the slice axis to smooth out the discontinuities.
Formulated in the gradient domain, this corresponds to finding the image whose intra-slices derivatives agree with
the intra-slice derivatives of the input data, but whose cross-slice derivatives are close to zero. Specifically,
letting $\Omega$ denote the voxel domain, $I^0$ denote the input voxel data, with slices stacked along the
$z$-axis, and $I^1$ denote the processed image, we define the target gradient field as:
$$\vec{G}=(\partial I^0/\partial x , \partial I^0/\partial y , 0)^t$$
and seek the output grid minimizing:
$$I^1 = \mathop{\min}_{I}\int_{\Omega}
\left(\nabla I-\vec{G}\right)^t W \left(\nabla I-\vec{G}\right)^t dp
$$
where $W=\hbox{Diag}(\beta_x,\beta_y,\beta_z)$ is a diagonal matrix with weights $\beta_x, \beta_y, \beta_z\geq0$
giving different importance to the different components of the gradients. For our application, we take
$\beta_x=\beta_y=1.0$ and $\beta_z=0.1$ to bias the processing in favor of preserving detail at the cost of less smoothing.
Using the Euler-Lagrange formulation, the minimizer of the (non-negative) quadratic energy is obtained by solving the anisotropic
diffusion problem:
\begin{equation}
\Delta_W I^1 = \nabla\cdot W \vec{G}
\label{eq:anisotropic_diffusion}
\end{equation}
where $\nabla$ is the gradient operator, $\nabla\cdot$ is the divergence operator, and $\Delta_W$ is the anisotropic Laplace operator
$\Delta_W = \nabla\cdot W \nabla$.

\subsubsection{Screened-Poisson Blending}
As visualized in Figure~\ref{f:data} (middle), though anisotropic diffusion effectively removes the inter-slice
discontinuities (top), it also blends out the details within each image (bottom). Specifically,
when considered on a slice-by-slice basis, the slices of $I^1$ have the correct low-frequency content, but lack
the high frequency content present in the slices of $I^0$. This motivates a second processing stage in which we
generate the slices of the new voxel grid, $I^2$, by using the low frequency data from the slices of $I^1$ and
high frequency data from the slices of $I^0$.

In the gradient domain, we seek a voxel grid whose $i$-th slice minimizes:
$$I^2_i = \mathop{\min}_I\int_{\Omega_i}
\alpha\left(I-I^1_i \right)^2 + \left\|\nabla I-\nabla I^0_i\right\|^2 dp$$
where $\Omega_i$ is the slice domain and $\alpha$ is the screening weight ($\alpha=0.01$ in our application) balancing the importance of interpolating pixel values of $I^1_i$ with the goal of matching the gradients of $I^0_i$. 
Using the Euler-Lagrange formulation, the minimizer is obtained by solving the screened Poisson equations:
$$(\alpha-\Delta) I^2_i = \alpha I^1_i - \Delta I^0_i$$
for each slice $i$. Alternatively, setting $\pi$ to be the projection onto the $xy$-plane, $\pi=\hbox{Diag}(1,1,0)$, the set of equations (across the different slices) can be consolidated into a single equation:
\begin{equation}
(\alpha-\Delta_\pi) I^2 = \alpha I^1 - \Delta_\pi I^0.
\label{eq:screened_poisson}
\end{equation}

As visualized in Figure~\ref{f:data} (right), the screening effectively combines the frequency data from the
two slices, providing a voxel grid that has the sharp intra-slice detail of the input without the inter-slice
discontinuities.

\subsection{Frequency-Space Interpretation}
Following the work of Bhat~{\em et al}.~\cite{Bhat:ECCV:2008}, we provide a frequency-space interpretation of the processing. To this end, we use the fact that the complex exponentials, $\phi_k(\theta)=e^{ik\theta}$, are eigenvalues of the differentiation operator, $\phi_k' = ik \phi_k$. And, more generally, if we denote by $\phi_{klm}(x,y,z)$ the three-dimensional complex exponential, $\phi_{klm}(x,y,z) = \phi_k(x)\cdot\phi_l(y)\cdot\phi_m(z)$, then:
$$\Delta_W(\phi_{klm}) = -\left(\beta_xk^2+\beta_yl^2+\beta_zm^2\right)\phi_{klm}.$$

\subsubsection{Anisotropic Diffusion}
Using $\hat{I}$ to denote the Fourier coefficients of signal $I$:
$$I(x,y,z)=\sum_{k,l,m}\hat{I}_{klm}\phi_{klm}(x,y,z)$$
we express the two sides of Equation~(\ref{eq:anisotropic_diffusion}) as:
\begin{align*}
\Delta_W I^1 &= -\sum_{k,l,m}\hat{I}^1_{klm}\left(\beta_xk^2+\beta_yl^2+\beta_z^2m^2\right)\phi_{klm}\\
\nabla\cdot W \vec{G} &= -\sum_{k,l,m}\hat{I}^0_{klm}\left(\beta_xk^2+\beta_yl^2\right)\phi_{klm}.
\end{align*}
Thus, the Fourier coefficients of the diffused signal can be written as:
$$\hat{I}^1_{klm} = \frac{\beta_xk^2+\beta_yl^2}{\beta_xk^2+\beta_yl^2+\beta_zm^2}\hat{I}^0_{klm}.$$

\subsubsection{Screened Poisson}
Similarly, we express the two sides of Equation~(\ref{eq:screened_poisson}) as:
\begin{align*}
(\alpha-\Delta_\pi)I^2 &= \sum_{k,l,m}\hat{I}^2_{klm}(\alpha+k^2+l^2)\phi_{klm}\\
\alpha I^1-\Delta_\pi I^0 &= \sum_{k,l,m}\left(\alpha\hat{I}^1_{klm}+(k^2+l^2)\hat{I}^0_{klm}\right)\phi_{klm}.
\end{align*}
Thus, the Fourier coefficients of the screened signal can be written as:
$$\hat{I}^2_{klm} = \frac{\alpha\hat{I}^1_{klm}+(k^2+l^2)\hat{I}^0_{klm}}{\alpha+k^2+l^2}.$$

\subsubsection{Combined System}
Thus, performing anisotropic diffusion followed by screening is equivalent convolving the input signal, $I^0$ with a filter $F$ whose coefficients are given by:
$$\hat{F}_{klm} = \frac{\alpha\frac{\beta_xk^2+\beta_yl^2}{\beta_xk^2+\beta_yl^2+\beta_zm^2}+(k^2+l^2)}{\alpha+k^2+l^2}.$$
Examining this filter, we observe:
\begin{itemize}
\item As $\beta_z\rightarrow0$ or $\alpha\rightarrow0$, we have $\hat{F}_{klm}\rightarrow1$. That is, if either there is no diffusion or there is no interpolation of the diffused signal, the signal is unchanged.
\item As $k^2+l^2\rightarrow\infty$, we have $\hat{F}_{klm}\rightarrow1$. That is, the filtering preserves large intra-slice frequencies.
\item As $k^2+l^2\rightarrow0$ and $m^2\rightarrow\infty$, we have $\hat{F}_{klm}\rightarrow0$. That is, the filtering dampens large inter-slice frequencies when the corresponding intra-slice frequency is low.
\end{itemize}

%% file: implementation.tex
The implementation of our 3D Poisson solver follows the earlier implementation of a 2D multigrid Poisson
solver for gradient-domain image processing~\cite{Kazhdan:SIGGRAPH:2008}. The solver is designed
to find the minimizer of quadratic energies of the form:
$$I^1 = \mathop{\min}_I\int_{\Omega}
\alpha\left(I-I^0\right)^2 + \left(\nabla I-\vec{G}\right)^tW\left(\nabla I-\vec{G}\right) dp$$
which can be obtained by solving:
$$(\alpha-\nabla\cdot W \nabla) I^1 = \alpha I^0 - \nabla\cdot W\vec{G}$$
where $I^1$ is the solution, $I^0$ prescribes the values, $\vec{G}$ prescribes
the gradients, $W$ is a non-negative diagonal matrix giving importance
to the individual gradient directions, and $\alpha$ is the screening weight.

\subsection{Multigrid Implementation}
Our multigrid implementation performs multiple iterations of a V-cycle solve. In the fine-to-coarse restriction phase
the solver performs Gauss-Seidel relaxation at the finest resolution, computes and down-samples the residual
to obtain the constraints for the next coarser level, and then repeats the process at the coarser level. At
the coarsest level a direct solver is used to solve the system. Then, in the coarse-to-fine prolongation phase,
the coarser solution is up-sampled and introduced as a correction term at the next (finer) resolution, and Gauss-Seidel
relaxations are applied once more before prolonging to the next finer resolution.

Each V-cycle is implemented as two streaming passes through the data. A window is maintained over the data, so that
(at each resolution) only a small subset of image slices resides in working memory and temporal
blocking~\cite{Pfeifer:CACM:1963,Douglas:ETNA:2000} is used to perform multiple Gauss-Seidel relaxations in a single streaming
pass. 

Specifically, each time the window is advanced, a new slice is read into the front of the window, the center slices are relaxed
in a front-to-back fashion, and the back slice is flushed to disk. We augment the traditional algorithm by performing multiple
relaxation passes over the center slices. We have found that this improves the solution without increasing either the I/O
(as would be required if we used more V-cycles) or the working memory size (as would be required if we
used temporal blocking with more Gauss-Seidel iterations).

\ignore
{
For the restriction phase, processing is performed by:
(1)~Advancing the window forward one slice;
(2)~Relaxing the voxels in a front-to-back order;
(3)~Computing the residual at the back of the window;
(4) Writing the solution and constraints at the back of the window to disk;
And, on every other iteration, (5)~Down sampling the computed residual slices to set the constraints for the
front of the window at the next coarser resolution.
For the finest resolution, advancing the window requires reading in the next slice of both the constraints and solution
from disk. For coarser resolutions, the solution is initialized to zero and the constraints are set through down-sampling. 

The prolongation phase is performed similarly only instead of computing and down-sampling the residual, the constraints
and previously computed solution are read from disk to the front of the window and the up-sampled solution is added to them.
}

\subsubsection*{Parallelization} We parallelize our implementation with OpenMP~\cite{openmp} using multi-coloring to parallelize the Gauss-Seidel relaxation
and accumulating coefficient values when performing the residual calculation, up-sampling, and down-sampling to avoid
write-on-write conflicts.

\subsubsection*{Disk I/O} For both the restriction and prolongation phases, disk I/O is performed using pre-fetching and lazy write-backs so as to
hide the I/O behind the computation instead of blocking while the data is read from and written to disk. Additionally,
though all computation is performed using single-precision floating point values, intermediate solutions and constraints
are stored on disk using half-precision floating point values~\cite{openexr} to reduce disk utilization. (Though this introduces a
small amount of high-frequency error between the restriction and prolongation phases, such error is quickly corrected
by Gauss-Seidel relaxation.)

\subsubsection*{Memory Usage} Our solver maintains a sliding window over both the constraints and the solutions.
When solving using $k$ Gauss-Seidel iterations, we store $k+2/3+2$ constraint slices in memory, corresponding to the
constraints for the $k$ slices being solved, two/three slices at the back of the window for storing the residual in the
restriction phase (depending on which of the two down-sampling operators described below we use), and additional front
and back slices for pre-fetching and lazy write-backs. We also store $k+2+1/2+2$ solution slices, corresponding to the
$k$ solutions slices being solved and their one-ring neighbors, one/two additional
slices at the front of the window for accumulating the correction term in the prolongation phase (depending on which of
the two up-sampling operators we use), and additional front and back slices for pre-fetching and lazy write-backs. In sum,
performing $k$ Gauss-Seidel iterations requires storing $2k+6+3/5$ slices in memory.

\begin{figure}[h]
\begin{center}
\ifarxivversion
\includegraphics[width=.9\columnwidth,natwidth=677,natheight=1025]{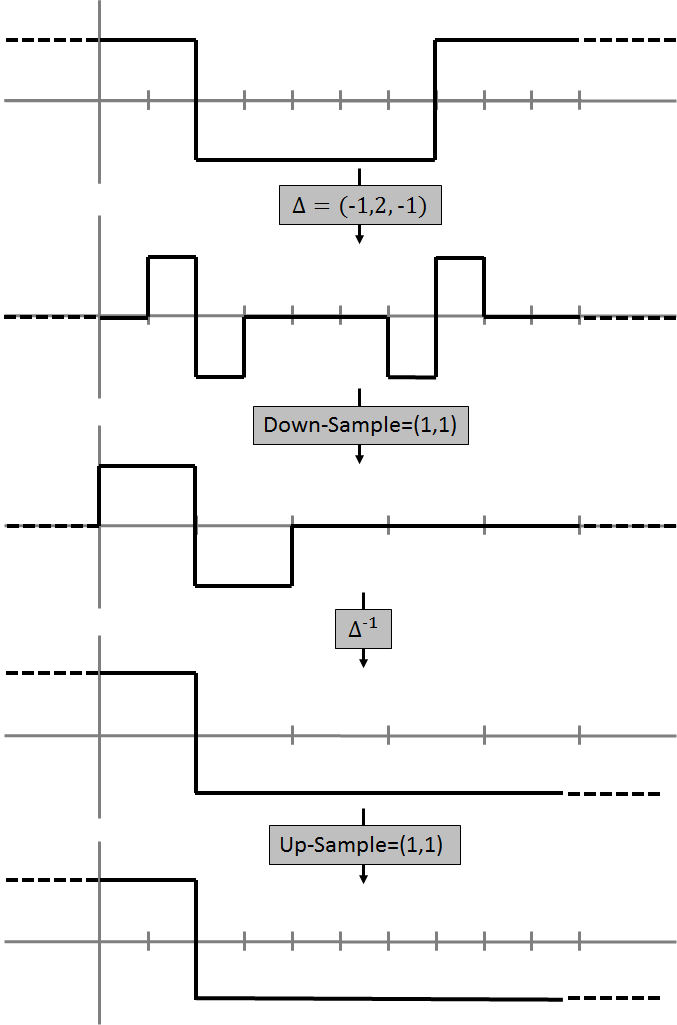}
\fi
\end{center}
\caption{
\label{f:down_sampling}
Aliasing artifacts introduced using the stencil $(1,1)$ for restriction/prolongation: Starting with the input signal (top), we compute the Laplacian constraints (top middle), down-sample the constraints to the coarser resolution (middle), solve the Poisson equation at the coarser resolution (bottom middle), and up-sample the solution to the finer level (bottom). After applying the Laplacian, the constraints have the typical $\hbox{+}1/\hbox{-}1$ and $\hbox{-}1/\hbox{+}1$ features around the sharp drops and rises of the original signal. Down-sampling with the $(1,1)$ stencil, the $\hbox{-}1/\hbox{+}1$ feature vanishes because the values fall into the same coarser cell and cancel each other out. As a result, the coarser level only ``sees'' the sharp drop in the signal, and the up-sampled solution does not reproduce the corresponding rise.
}
\end{figure}

\subsection{Discretization}
To define the multigrid system, we need both a discretization of the Laplacian and a choice of prolongation operator
to up-sample solutions from coarser levels to finer ones. We consider two common approaches: ({\bf Constant})~using the 6-point
Laplacian stencil (defined by setting matrix coefficients of face-adjacent neighbors to -1) with a prolongation
operator that replicates the solution into the eight children (using the stencil $(1,1)^{\otimes3}$), and ({\bf Linear})~using
the 26-point Laplacian stencil defined by using first-order B-splines as finite elements with the prolongation
operator defined by the nesting of B-splines (using the stencil $(1/2,1,1/2)^{\otimes3}$).

The advantage of the first approach is that a voxel's value is only linked to its six neighbors, so Gauss-Seidel
relaxation is computationally inexpensive. However, the prolongation operators is not as smooth, and high-frequencies
can be introduced in the course of the up-/down-sampling, resulting in a less efficient multigrid. (For a simple example, see Figure~\ref{f:down_sampling}.)
In contrast, the implementation based on first-order B-splines results in more expensive Gauss-Seidel relaxation but gives rise to
a more efficient multigrid solver.

In addition to the two common approaches, we have also explored an approach that tries to capture the best of both methods  ({\bf Hybrid}). We use the 6-point Laplacian
at the finest resolution but use the prolongation operator for the first-order B-splines to transition between the
levels of the hierarchy. Although using the Galerkin formulation~\cite{Fletcher:Springer:1984} results in a 26-point
Laplacian at coarser resolutions, each coarser resolution is 8 times smaller and we do not expect the additional computational complexity of relaxing
at coarser resolutions to contribute significantly to the overall running time.

%% file: results.tex
To evaluate our solver, we ran both the anisotropic diffusion and the screened-Poisson blending on the $21496\times25792\times1850$ Mouse S1 dataset~\cite{Kat11}, imaging a cortical region at $3\times3\times30$ $\mu m^3$ spatial resolution, to remove the inter-slice variation
\ifarxivversion
(Figure~\ref{f:data}).
\fi
The performance of our solver is summarized in Table~\ref{t:stats}. For the 3D anisotropic diffusion we used two V-cycles with three relaxation passes per cycle and three Gauss-Seidel iterations per pass. Since the 2D anisotropic diffusion is performed one slice at a time and the slices align with the stream order, memory usage is less restrictive and we used a single V-cycle with one relaxation per pass and ten Gauss-Seidel iterations.

Looking at the results in Table~\ref{t:stats}, we make several observations:
(1)~Though {\bf Linear} gives the smallest residual, it is also significantly slower than the other two methods because 3D {\bf Linear} couples a voxel's value to its 26 (8 for 2D systems) neighbors  while 3D {\bf Constant} and {\bf Hybrid} couple the value to only the 6 (4 for 2D systems) neighbors (at the finest resolution). Note that {\bf Hybrid} is slightly slower than {\bf Constant} because its down-sampling requires accumulating from $3^3$ values rather than $2^3$ as in {\bf Constant}.
(2)~All three discretizations have roughly the same memory usage. ({\bf Linear} and {\bf Hybrid} use slightly more memory because they require buffering an additional slice for both the constraints and the solution.)
(3)~Running times and memory are significantly smaller for the screened-Poisson blending since it is formulated as a set of 2D linear systems, which can be implemented in a single streaming pass that only requires a window of one slice to be in the working memory at any given time. Thus, the implementation uses less memory and does not incur the additional cost of reading/writing the constraints and solution from/to disk.

\begin{table}[h]
\center{
\begin{tabular}{l|c|c|c}
& Time & Memory & Residual \\
& {\scriptsize(AD/SP)} & {\scriptsize(AD/SP)} & {\scriptsize(AD/SP)} \\
\hline
{\bf Constant} & 40:30 / 16:24 & 42 / 31 & $4.7\!\!\times\!\!10^{\hbox{-}4}$ / $4.7\!\!\times\!\!10^{\hbox{-}4}$ \\
{\bf Hybrid}   & 41:16 / 17:03 & 45 / 31 & $2.6\!\!\times\!\!10^{\hbox{-}4}$ / $4.5\!\!\times\!\!10^{\hbox{-}4}$ \\
{\bf Linear}   & 80:41 / 23:09 & 45 / 31 & $6.1\!\!\times\!\!10^{\hbox{-}5}$ / $5.0\!\!\times\!\!10^{\hbox{-}4}$ \\
\hline
\end{tabular}
}
\caption{
\label{t:stats}
Solver Performance: Times are given as {\em hours}:{\em minutes}, memory is measured in gigabytes, and residual is measured as the ratio of the $L_2$-norm of the residual and the $L_2$-norm of the initial constraints. Results are shown separately for the anisotropic diffusion (AD) and screened Poisson (SP) phases.
}
\end{table}

\begin{figure}[h]
\begin{center}
\ifarxivversion
\includegraphics[width=\columnwidth,natwidth=682,natheight=290]{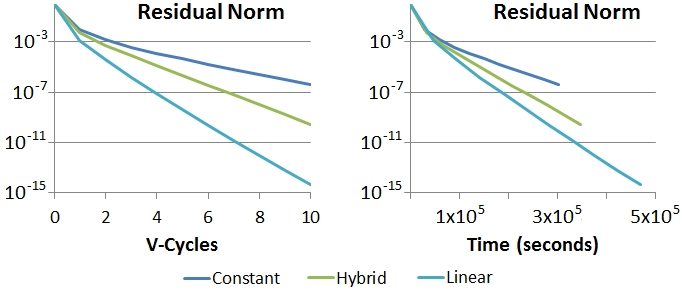}
\fi
\end{center}
\caption{
\label{f:timing}
Residual norm reduction as a function of the number of V-cycles (left) and running time (right).
}
\end{figure}

To better assess the relative benefits of the three solvers, we re-ran the anisotropic diffusion experiment using double-precision floating  points for both in-memory and on-disk storage, and we measured the decrease in residual norm over 10 V-cycles.\footnote{Due to the increased storage size, these experiments were performed on a down-sampled, $10748\times12896\times1850$, version of the dataset.} The results of these experiments can bee seen in Figure~\ref{f:timing} which plots the residual norm as a function of V-cycles (left) and running time (right).

As the plots on the left indicate, all three implementations exhibit standard multigrid behavior, with residual norms decaying exponentially in the number of V-cycles performed. Furthermore, as we would expect, the higher-order {\bf Linear} solver exhibits better convergence per V-cycle (with respective decay rates of $0.23$, $0.11$, and $0.04$ per V-cycle for {\bf Constant}, {\bf Hybrid}, and {\bf Linear}). However, when taking the running time into account, the discrepenancy between {\bf Hybrid} and {\bf Linear} becomes significantly less prononouced (with respective decay rates of $0.62$, $0.53$, and $0.50$ per $10^5$ seconds for {\bf Constant}, {\bf Hybrid}, and {\bf Linear}).

%% file: discussion.tex
In this work, we have focused on extending the established 2D gradient-domain image processing paradigm to the context of processing 3D voxels. As our work has sought to address specific imaging artifacts that arise in the context large EM stacks, we have focused on the realization of a particular filter. However, we stress that the solver we have described supports general purpose gradient-domain filtering, making it possible to apply the variety of filters described in~\cite{Bhat:SIGGRAPH:2010} to the processing of large, out-of-core, voxel grids, including applications such as HDR compression, gradient amplification, smoothing, and non-photorealistic filtering.

\begin{figure*}[ht]
\begin{center}
\ifarxivversion
\includegraphics[width=.67\columnwidth,natwidth=960,natheight=1080]{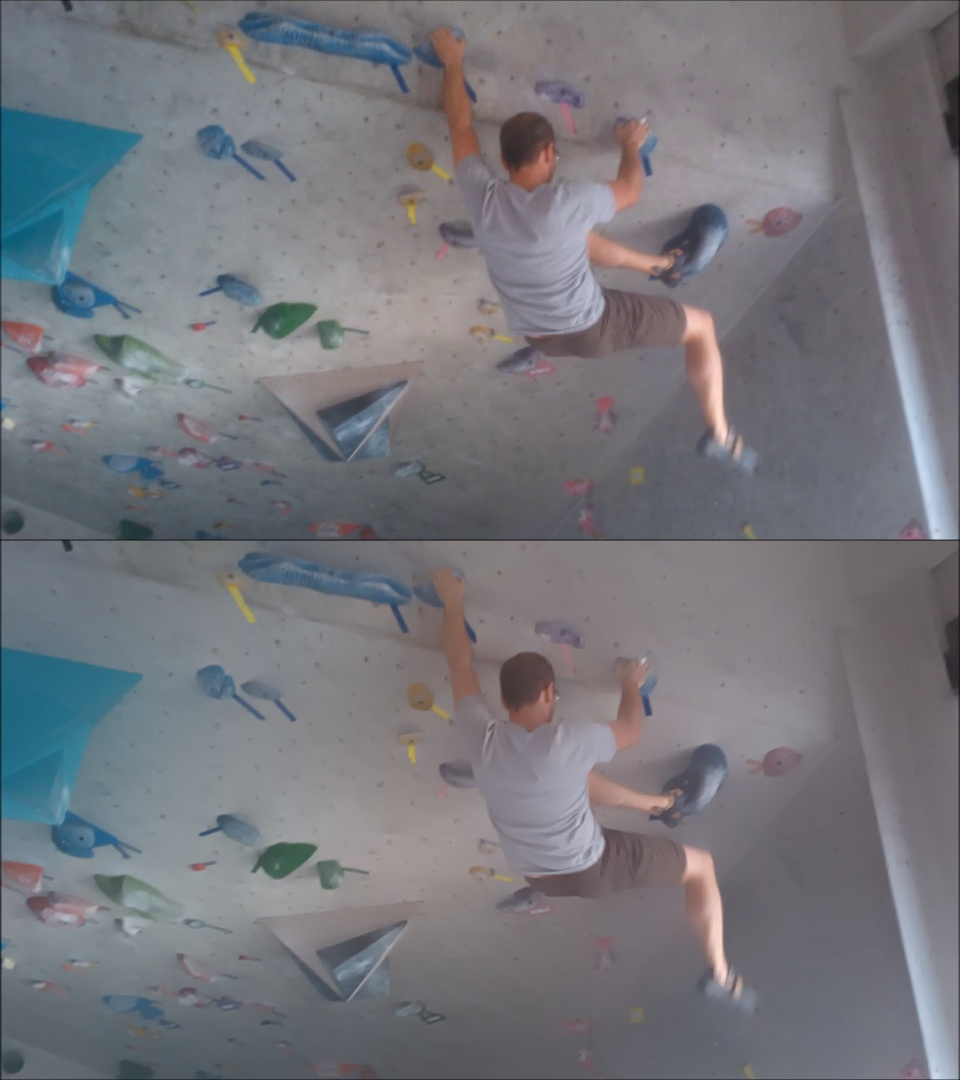}
\includegraphics[width=.67\columnwidth,natwidth=960,natheight=1080]{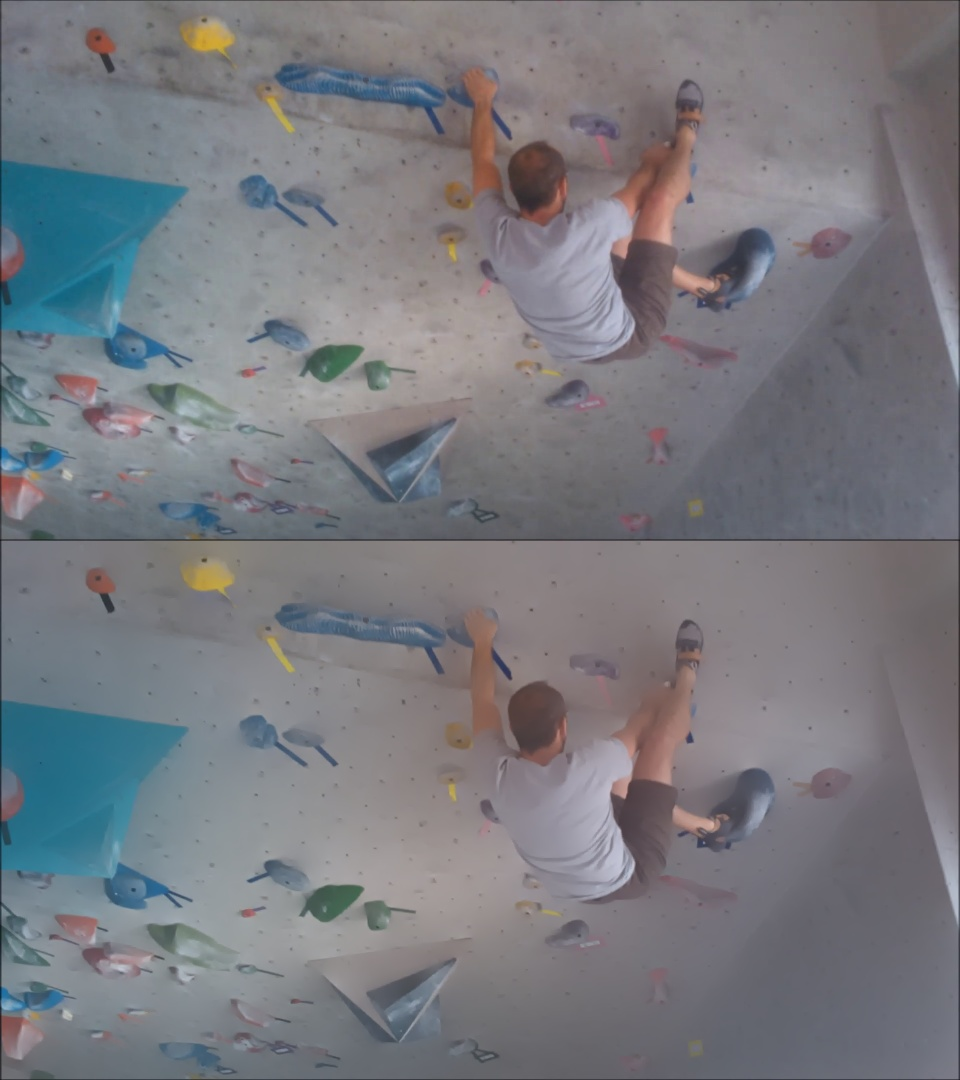}
\includegraphics[width=.67\columnwidth,natwidth=960,natheight=1080]{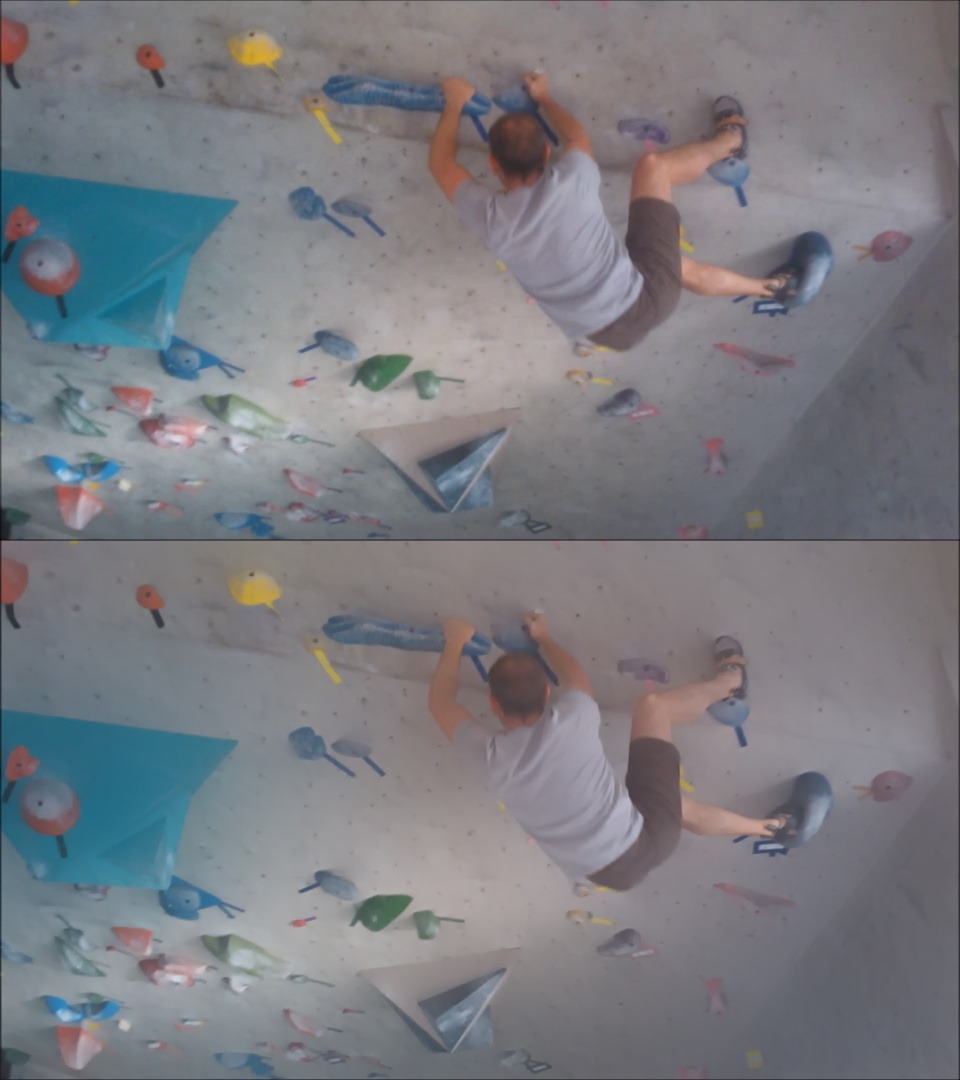}
\fi
\end{center}
\caption{
\label{f:video_processing}
Applying a non-linear filtering to a $1920\times1080\times742$ video volume: The images on top shows adjacent frames in the input video. The images on the bottom show the same frames after solving a Poisson equation to supress small gradients.
}
\end{figure*}

As an example, Figure~\ref{f:video_processing}
shows an application of our solver to non-linear, out-of-core, processing of a $1920\times1080\times742$ video -- reducing noise in the video by suppressing small gradients. The output, $I^1$, was obtained by solving a global screened Poisson equation:
$$I^1 = \min_{I}\int_\Omega \alpha(I-I^0)^2 + \left\|\nabla I - \vec{V}_{\lambda,\sigma}\right\|_W^2$$
where the vector field $\vec{V}_{\lambda,\sigma}$ is obtained by modulating the gradients of the input as a function of their magnitude, following the NPR filter proposed by Bhat~{\em et al.}~\cite{Bhat:SIGGRAPH:2010} (Section~6.2):
$$\vec{V}_{\lambda,\sigma} = \lambda\left(1-\exp\left(\frac{-\|\nabla I^0\|^2}{2\sigma^2}\right)\right)\nabla I^0.$$
Here, we set the screening weight to $\alpha=0.001$, the gradient weight tensor to $W=\hbox{Diag}(1,1,0.1)$ and the amplitude and standard deviation of the gradient clamping parameters to $\lambda=1.25$ and $\sigma=5$, (with gray-scale values in the range $[0,256)$).

Using our approach, the processing ran with a peak memory footprint of 185~MB per channel. By comparison, the total memory for representing the constraints and solution with single floating point precision would be almost two orders of magnitude larger, requiring 11~GB per channel.

%% file: conclusion.tex
To support the broader class of inhomogeneous gradient-domain techniques, we would like to extend the method to support spatially varying weights for both the screening and gradient terms. This would make it possible to use our solver to perform tasks like edge-aware diffusion for facilitating anatomy segmentation. We would also like to extend our implementation to support distributed processing as in~\cite{Kazhdan:TOG:2010} in order to make our solver more efficient and reduce the memory load on individual machines.